\documentclass[aps,prd,superscriptaddress,preprintnumbers]{revtex4-2}
\usepackage{graphicx,float,wrapfig,subfigure}
\usepackage{amsfonts,amsmath,amssymb,amstext}
\usepackage{latexsym}
\usepackage{bm}
\usepackage{color}

\usepackage{slashed}
\usepackage[normalem]{ulem}

\usepackage{enumitem}
\usepackage{placeins}

\newcommand{\be}{\begin{equation}}
\newcommand{\ee}{\end{equation}}
\newcommand{\ba}{\begin{eqnarray}}
\newcommand{\ea}{\end{eqnarray}}

\definecolor{red}{rgb}{0.7,0,0}
\definecolor{RED}{rgb}{0.7,0,0}
\definecolor{green}{rgb}{0,0.5,0}

\begin{document}

\title{Determining the NJL Coupling and AMM in Magnetized QCD Matter via Machine Learning}
\date{\today}
\author{Zigeng Ding}
\author{Fan Lin}
\email{linfan@aust.edu.cn}
\author{Xinyang Wang}
\email{wangxy@aust.edu.cn, corresponding author}
\affiliation{Center for Fundamental Physics, School of Mechanics and Physics, Anhui University of Science and Technology, Huainan, Anhui 232001, People's Republic of China}

\begin{abstract}
	In this study, we investigate the phase structure of magnetized QCD matter by determining the field-dependent parameters of the Nambu-Jona-Lasinio (NJL) model through a physics-informed machine learning framework. Specifically, we focus on extracting the optimal functional forms for the running coupling constant $G(eB)$ and the quark anomalous magnetic moment (AMM) ratio $v_2(eB)$, using lattice QCD computation on the chiral transition and condensate behavior. By embedding the NJL gap equation as a differentiable physics-constrained module, our neural network pipeline identifies continuous parameter functions that accurately reproduce the inverse magnetic catalysis (IMC) effect. Our results demonstrate that the magnetic field smoothly suppresses both $G$ and $v_2$. This approach not only bridges the gap between effective models and lattice data but also provides new microscopic insights into the response of the QCD vacuum to strong magnetic fields.  As an independent test, the neutral-pion mass is calculated with the mean learned parameters and reproduces the decreasing trend of recent lattice results without a pion-data refit.
\end{abstract}
\maketitle
\section{Introduction}

Ultra-strong magnetic fields arise in a variety of extreme environments, ranging from the transient fields produced in relativistic heavy-ion collisions to the enormous surface fields of magnetars. Such intense magnetic backgrounds provide a unique laboratory for exploring strongly interacting matter, as they can significantly modify the structure of the quantum chromodynamics (QCD) vacuum and influence the properties of hadronic and quark matter. Understanding how QCD behaves under these conditions has therefore become an important problem in nuclear and high-energy physics. In particular, the phase structure of QCD matter in strong magnetic fields remains a central puzzle, closely related to fundamental issues such as chiral symmetry breaking and modifications to the QCD phase diagram.

Early theoretical work predicted that the chiral critical temperature would rise with increasing magnetic field—a phenomenon known as magnetic catalysis (MC)~\cite{Klevansky:1989vi,Klimenko:1990rh,Gusynin:1995nb,Shovkovy:2012zn}. However, subsequent lattice QCD studies revealed a strikingly different behavior around the transition region: the critical temperature for chiral symmetry restoration decreases with the magnetic field, an effect termed inverse magnetic catalysis (IMC)~\cite{Bali:2011qj,Bali:2012zg,Ding:2022tqn}. This counterintuitive phenomenon has attracted significant theoretical interest, prompting numerous investigations into its underlying physical origin, which remains incompletely understood.

The Nambu–Jona-Lasinio (NJL) model~\cite{Nambu:1961tp,Nambu:1961fr}, as a successful low-energy effective theory, provides a powerful and tractable framework for describing chiral symmetry breaking and meson physics. In the conventional treatment, the model’s coupling constant is held fixed, which invariably leads to magnetic catalysis~\cite{Shovkovy:2012zn}. Recent studies have proposed a more realistic approach by promoting the four-fermion coupling $G$ to a magnetic-field-dependent quantity $G(eB)$. By calibrating $G(eB)$ against lattice QCD data, the IMC effect can be naturally reproduced~\cite{Farias:2014eca,Ferreira:2014kpa,Liu:2016vuw,Sheng:2021evj,Avancini:2021pmi,Mao:2022nfs,Mao:2024gox,Mao:2025toi}. This approach reflects a physically motivated picture: strong magnetic fields inevitably modify non-perturbative QCD dynamics and hadron structure. While first-principle calculations in this regime are challenging, effective models such as the NJL model offer a natural setting in which the effective coupling can “run” with the field strength, thereby encoding vacuum polarization and other non-perturbative responses to the magnetic field.

Furthermore, the quark anomalous magnetic moment (AMM), another manifestation of chiral symmetry breaking, has attracted considerable attention~\cite{Ferrer:2014qka,Ferrer:2009nq,Ferrer:2013noa, Bicudo:1998qb,Chang:2010hb, Xing:2021kbw, Mao:2018jdo, Mei:2020jzn, Chaudhuri:2019lbw, Ghosh:2020xwp, Strickland:2012vu, Chaudhuri:2020lga, Ferrer:2015wca, Fayazbakhsh:2014mca, Chao:2020jjy, Farias:2021fci}. Previous studies show that the quark AMM suppresses the chiral condensate in both zero and finite-temperature regimes. Additionally, the quark AMM is an intrinsic physical parameter that becomes non-negligible in the presence of strong magnetic fields. Although the microscopic origin of the AMM is a complex problem, recent studies—borrowing from QED estimates where the lepton AMM is proportional to the lepton mass~\cite{Lin:2021bqv}—favor a magnetic-field-dependent form proportional to the square of the scalar mean field, $\sigma^2$~\cite{Lin:2022ied, Kawaguchi:2022dbq, Wei:2022oyb, Kawaguchi:2024edu}. In most existing works, the proportionality ratio is treated as a constant, with field dependence entering only through the scalar mean field. More generally, however, this ratio may also depend on the magnetic field and should be taken into account.

In recent years, machine learning, particularly neural networks, has demonstrated immense potential for discovering physical laws from high-dimensional data and constructing constitutive relations~\cite{Schmidhuber:2014bpo,LeCun:2015pmr,Boehnlein:2021eym, Larkoski:2019yxx, Baldi:2016fgs, Du:2020pmp, Pang:2016vdc, Pang:2019aqb, Du:2019civ, He:2023tvu, Zhou:2023pti, Ma:2023zfj, Wang:2023kcg, He:2023urp, Wang:2020tgb, Huang:2025uvc,Guo:2023mhf,Chen:2024mmd,Chen:2024ckb,Zhang:2026zoz}. It offers a new paradigm to bridge the gap between lattice QCD data and effective models. The core idea of this work is to employ machine learning as a powerful tool for inverse engineering, using lattice QCD computation on the chiral transition and condensate behavior and a parameterized (field-dependent) NJL model as the ``interpretable function". The transition-temperature intervals and reference constituent masses provide numerical targets, while condensate trends enter as physical priors. We train a neural network to directly ``learn" the optimal functional forms for the evolving coupling constant $G(eB)$ and AMM ratio $v_2(eB)$ which is defined by $\kappa(eB) = v_2(eB) \sigma^2$.

This paper aims to develop and apply this framework. We first construct a lattice-improved NJL model incorporating a magnetic-field-dependent coupling and AMM. Subsequently, we design a supervised learning pipeline: using the forward NJL model calculation as a differentiable physics-constrained module and targeting lattice QCD data, we employ automatic differentiation and gradient descent optimization to invert for a set of continuous and flexible $G(eB)$ and $v_2(eB)$ functions constrained by the selected inputs. This approach avoids imposing a fixed analytic form for the running parameters and can reveal complex evolutionary patterns difficult to detect with traditional fitting methods. Ultimately, the learned parameter evolution will provide new microscopic insights into the response of the QCD vacuum to magnetic fields and yield a more precise and reliable parameterized effective model for studying the QCD phase diagram and related phenomena.

The paper is structured as follows: Section~\ref{sec:framework} reviews the theoretical framework of the NJL model in a magnetic field and the introduction of running parameters. Section~\ref{sec:method} details our machine learning methodology and model architecture. Section~\ref{sec:results} presents the results for $G(eB)$ and $v_2(eB)$ learned from lattice data, along with physical analysis. Section~\ref{sec:pion_results} tests the learned parameters using the neutral-pion mass. Section~\ref{sec:conclusion} provides a summary and outlook for future work.

\section{Theoretical Framework: NJL Model with Field-Dependent Parameters}
\label{sec:framework}

The Lagrangian density of the two-flavor Nambu--Jona-Lasinio (NJL) model in the presence of a background magnetic field, including the contribution from the anomalous magnetic moment (AMM) of quarks, is given by~\cite{Fayazbakhsh:2014mca}
\begin{equation}
	\mathcal{L}
	=\bar{\psi}\left(i\gamma^{\mu}D_{\mu}-\hat{m}
	+\frac{1}{2} q_f \kappa F_{\mu\nu}\sigma^{\mu\nu}\right)\psi
	+G \left[(\bar{\psi}\psi)^2+(\bar{\psi}i\gamma^5\boldsymbol{\tau}\psi)^2\right],
\end{equation}
where $\hat{m}=\mathrm{diag}(m_u,m_d)$ is the current quark mass matrix,
$q_f=\mathrm{diag}(2e/3,-e/3)$ denotes the quark charge matrix, and
$\kappa=\mathrm{diag}(\kappa_u,\kappa_d)$ represents the AMM matrix.
Here $e$ denotes the absolute value of the electron charge.
The covariant derivative
\begin{equation}
	D_\mu=\partial_\mu-i q_f A_\mu,
\end{equation}
incorporates the coupling between quarks and the external electromagnetic field.

To determine the dynamical quark mass, we introduce a homogeneous scalar auxiliary field $\sigma$ in the mean-field approximation and restrict the calculation to a vanishing pseudoscalar background. At this stage, the AMM matrix $\kappa=\mathrm{diag}(\kappa_u,\kappa_d)$ is kept general, without specifying its functional form or imposing flavor independence. The resulting mean-field effective Lagrangian is
\begin{equation}
		\mathcal{L}_{\mathrm{MF}}
		=\bar{\psi}\left(i\gamma^{\mu}D_{\mu}-\hat m-\sigma
		+\frac12q_f\kappa F_{\mu\nu}\sigma^{\mu\nu}\right)\psi
		-\frac{\sigma^2}{4G}.
\end{equation}
where the constituent quark mass matrix is defined as
\begin{equation}
	\label{eq:mass_field}
	\hat M=\hat m+\sigma,
\end{equation}
with $\hat M=\mathrm{diag}(M_u,M_d)$ the unit flavor matrix implicit in the scalar term. Here $\sigma$ is the scalar self-energy. Its relation to the mass-source condensate follows from the full stationarity condition. Any prescribed dependence of $\kappa_f$ on $\sigma$ must be included when varying the potential; for independent AMM parameters, $\partial\kappa_f/\partial\sigma=0$.

The presence of the AMM modifies the Dirac equation for quarks to
\begin{equation}
	\left(
	i\gamma^\mu D_\mu
	-\hat{M}
	+\frac{1}{2}\kappa q_f \sigma^{\mu\nu}F_{\mu\nu}
	\right)\psi=0 .
\end{equation}
Consequently, the quark energy spectrum is also modified.
Following Ref.~\cite{Lin:2022ied}, we consider a uniform magnetic field
$\mathbf{B}=B\hat{z}, B\ge0$ directed along the $z$-axis.
The corresponding Landau-level dispersion relation for quarks becomes
\begin{equation}
	\omega_{f,l,s}^{2}=p_z^2+
		\left[\sqrt{M_f^2+2l|q_f B|}-s\kappa_f q_f B\right]^2.
\end{equation}
Here $l=0,1,2,\ldots$ labels the physical Landau levels. The lowest level $l=0$ contains only $s=\mathrm{sign}(q_f)$, whereas $l\ge1$ contains $s=\pm1$. All sums over $(l,s)$ below follow this counting, without an additional spin-degeneracy factor.

Within the mean-field approximation, the thermodynamic potential $\Omega$ can be obtained following the standard procedure described in Ref.~\cite{Buballa:2003qv}. It takes the form
\begin{equation}
	\Omega
	=
	\frac{\sigma^{2}}{4G}
	-
	N_c\sum_{f,l,s}
	\frac{|q_f B|}{2\pi}
	\int_{-\infty}^{+\infty}\frac{\mathrm{d}p_z}{2\pi}\,
	\omega_{f,l,s}
	-
	N_c\sum_{f,l,s}
	\frac{|q_f B|}{2\pi}
	\int_{-\infty}^{+\infty}\frac{\mathrm{d}p_z}{2\pi}\,
	2T\ln\!\left(1+e^{-\beta\omega_{f,l,s}}\right),
\end{equation}
where the first term represents the scalar mean-field contribution, while the second and third terms correspond to the vacuum and thermal contributions of quarks in the presence of the magnetic field. The regularization specified below is understood in both quark contributions.

The equilibrium scalar field is determined by minimizing the thermodynamic potential.  This condition is equivalent to solving the gap equation
\begin{equation}
	\label{eq:gap}
	\frac{\partial \Omega}{\partial \sigma}
	=
	\frac{\sigma}{2G}
	-
	N_c\sum_{f,l,s}
	\frac{|q_f B|}{2\pi}
	\int_{-\infty}^{+\infty}
	\frac{\mathrm{d}p_z}{2\pi}
	\left[
		1-2\left(1+e^{\beta\omega_{f,l,s}}\right)^{-1}
		\right]
	\frac{\partial \omega_{f,l,s}}{\partial \sigma}
	=0,
	\qquad
	\frac{\partial^2 \Omega}{\partial \sigma^2}>0 ,
\end{equation}
where $T=1/\beta$ denotes the temperature. The derivative includes any prescribed $\sigma$ dependence of $\kappa_f$, with $T$, $B$, $G$, the current masses, and the independent parameters specifying $\kappa_f$ held fixed. Positive curvature selects local minima along the scalar direction; the equilibrium branch is the lowest of these in the same regularized potential.

For clarity, defining $r_{fl}=\sqrt{M_f^2+2l|q_fB|}$ and $A_{fls}=r_{fl}-s\kappa_f q_fB$, the energy derivative entering Eq.~\eqref{eq:gap} is
\begin{equation}
	\label{eq:energy_derivative}
	\frac{\partial\omega_{f,l,s}}{\partial\sigma}
		=\frac{A_{fls}}{\omega_{f,l,s}}
		\left(\frac{M_f}{r_{fl}}-s q_fB\frac{\partial\kappa_f}{\partial\sigma}\right).
\end{equation}
The scalar condensate is instead defined by the current-mass source. For $m_u=m_d\equiv m$, let $C=\langle\bar uu+\bar dd\rangle$ denote the two-flavor sum, in the convention where it is negative in the vacuum. To isolate this operator, the source derivative is taken at fixed independent AMM coefficients, before imposing any relation between $\kappa_f$ and the current mass:
\begin{equation}
	\label{eq:source_condensate}
	\begin{aligned}
			C(B,T) & =\left.\frac{\partial\Omega}{\partial m}\right|_{T,B,G,\sigma,\{\kappa_f\}}
			\bigg|_{\sigma=\sigma_{\mathrm{eq}}}
			\\
			       & =-\sum_R\left[1-2n_F(\omega_{f,l,s})\right]
			\frac{A_{fls}}{\omega_{f,l,s}}\frac{M_f}{r_{fl}},
		\end{aligned}
\end{equation}
where $n_F(\omega)=(1+e^{\beta\omega})^{-1}$ and $\sum_R$ denotes the sum-integral $N_c\sum_{f,l,s}|q_fB|/(2\pi)\int\mathrm{d}p_z/(2\pi)$ with the Landau-level counting above and the regularization prescription below understood. Writing $\sum_{R,f}$ for the corresponding sum-integral at fixed flavor, we define the AMM responses
\begin{equation}
	\label{eq:amm_response}
	J_f(B,T)\equiv\left.\frac{\partial\Omega_q}{\partial\kappa_f}\right|_{\{M_g\},\{\kappa_{g\ne f}\}}
		=\sum_{R,f}\left[1-2n_F(\omega_{f,l,s})\right]
		\frac{A_{fls}}{\omega_{f,l,s}}\,s q_fB,
		\qquad J_2\equiv\sum_f J_f.
\end{equation}
where $\Omega_q$ is the quark part of $\Omega$ and the derivative treats the AMM coefficients as independent. The gap equation can equivalently be written as
\begin{equation}
	\label{eq:gap_source}
	\frac{\sigma}{2G}+C+\sum_f\frac{\partial\kappa_f}{\partial\sigma}J_f=0,
		\qquad M=m-2G\left(C+\sum_f\frac{\partial\kappa_f}{\partial\sigma}J_f\right).
\end{equation}
Thus $-\sigma/(2G)$ equals the mass-source condensate only when $\sum_f(\partial\kappa_f/\partial\sigma)J_f=0$, for example at $B=0$ or for $\sigma$-independent AMM coefficients. If a chosen parametrization also has explicit current-mass dependence, the equilibrium potential instead obeys $d\Omega_{\mathrm{eq}}/dm=C+\sum_f(\partial\kappa_f/\partial m)_{\sigma}J_f$ at fixed $T$, $B$, $G$, and independent model parameters. This total derivative includes an AMM-source contribution and must be distinguished from the scalar operator in Eq.~\eqref{eq:source_condensate}. No functional form for $\kappa_f$ is required for these identities. Changing the observable definition alone leaves the stationary solutions of a fixed, consistently implemented potential unchanged.

The identification $\langle\bar uu+\bar dd\rangle=-\sigma/(2G_S)$ is used in the Gell-Mann--Oakes--Renner analysis of Ref.~\cite{Lin:2022ied} (Sec.~III D), although its gap equation retains $\kappa'(\sigma)$. Under the source prescription of Eq.~\eqref{eq:source_condensate}, this identification omits the AMM response in Eq.~\eqref{eq:gap_source} and does not, in general, give the mass-source condensate when $\kappa$ depends on $\sigma$ but not explicitly on the bare mass. This issue is absent when the response vanishes. The source-derivative definition is also stated explicitly in Eq.~(7) of Ref.~\cite{Kawaguchi:2022dbq}.

A key unknown ingredient in the present framework is the functional form of the quark anomalous magnetic moment. In a vacuum, the AMM of a charged particle is generally proportional to the square of its mass. For quarks, however, the perturbative contribution is extremely small and therefore produces only negligible effects.

Recent studies based on the Dyson--Schwinger equations indicate that dynamical chiral symmetry breaking can generate a sizable AMM through the chiral condensate~\cite{Chang:2010hb}. Motivated by this observation, a nonperturbative ansatz for the quark AMM is commonly adopted in the form $\kappa = v_2\,\sigma^2,$
which incorporates the dynamical origin of the AMM. Here, $v_2$ is a phenomenological parameter characterizing the strength of this effect. In the present work, we further generalize this ansatz. Since the AMM arises dynamically from the QCD medium, it is natural to expect that the coefficient $v_2$ may also depend on the external magnetic field. We therefore allow $v_2 = v_2(eB)$ rather than treating it as a constant parameter. For the calculations below, we specialize the general expressions above to $\kappa_u=\kappa_d=v_2(eB)\sigma^2$. At fixed $v_2$, this gives $\partial\kappa_f/\partial\sigma=2v_2\sigma$ and $(\partial\kappa_f/\partial m)_\sigma=0$, so Eq.~\eqref{eq:gap_source} reduces to
	\begin{equation}
		\frac{\sigma}{2G}+C+2v_2\sigma J_2=0,
	\end{equation}
	with $C=d\Omega_{\mathrm{eq}}/dm$.

The NJL model is a nonrenormalizable effective theory and therefore exhibits ultraviolet divergences. To preserve its validity as a low-energy effective description, a regularization procedure must be introduced. In the present work, we adopt a three-dimensional soft-cutoff scheme to regularize the momentum integrals that diverge. Within this prescription, any integral requiring regularization is implemented as
\begin{equation}
	\label{eq:regularization_rule}
	\int \mathrm{d}k_z\, g(k_z)
	\;\longrightarrow\;
	\int \mathrm{d}k_z\, g(k_z)\, f_{\Lambda,q_f B}^{k_z},
\end{equation}
where $g(k_z)$ denotes an arbitrary function of the longitudinal momentum $k_z$. The regulator function is chosen as
\begin{equation}
	\label{eq:regulator}
	f_{\Lambda,q_f B}^{k_z}
	=
	\frac{\Lambda^{10}}{\Lambda^{10}+\left[k_z^{2}+2l|q_f B|\right]^5},
\end{equation}
with $\Lambda$ representing the ultraviolet cutoff scale.
Because the primary focus of this work is the quark chiral condensate behaviour, we apply the regularization directly to the gap equation~\eqref{eq:gap} as a whole, following the approach of Ref.~\cite{Xue:2021ldz}.
The model parameters in vacuum are fixed as follows: $G = 4.79877~\mathrm{GeV}^{-2}$, $\Lambda = 0.63704~\mathrm{GeV}$, and $m_u=m_d\equiv m = 5~\mathrm{MeV}$.
\begin{table*}
	\begin{ruledtabular}
		\begin{tabular}{lcccccccc}
			$\text{eB}~[\text{GeV}^2]$ & 0.0   & 0.1   & 0.2   & 0.3   & 0.4   & 0.5   & 0.6   & 0.7   \\
			\colrule
			$\text M^2~[\text{GeV}^2]$ & 0.097 & 0.096 & 0.094 & 0.091 & 0.087 & 0.083 & 0.079 & 0.074 \\
		\end{tabular}
	\end{ruledtabular}
	\caption{The values of the magnetic field and the corresponding constituent quark masses from lattice result~\cite{Bali:2012zg}.}
	\label{Table1}
\end{table*}

The lattice results for constituent quark masses under various magnetic field strengths $eB$ are summarized in Table~\ref{Table1}. Together with the (pseudo)phase transition point obtained from lattice simulations, these data allow us to determine the magnetic field-dependent parameters $G(eB)$ and $v_2(eB)$ through the gap equation~\eqref{eq:gap}. However, this task poses a significant challenge for conventional numerical methods, as we are faced with two unknown parameters but only one equation supplemented by a single constraint. This necessitates a machine-learning approach to solve the parameter-inversion problem effectively.

\section{Machine Learning Methodology}
\label{sec:method}

\begin{figure}[t]
	\centering
	\includegraphics[width=0.9\textwidth]{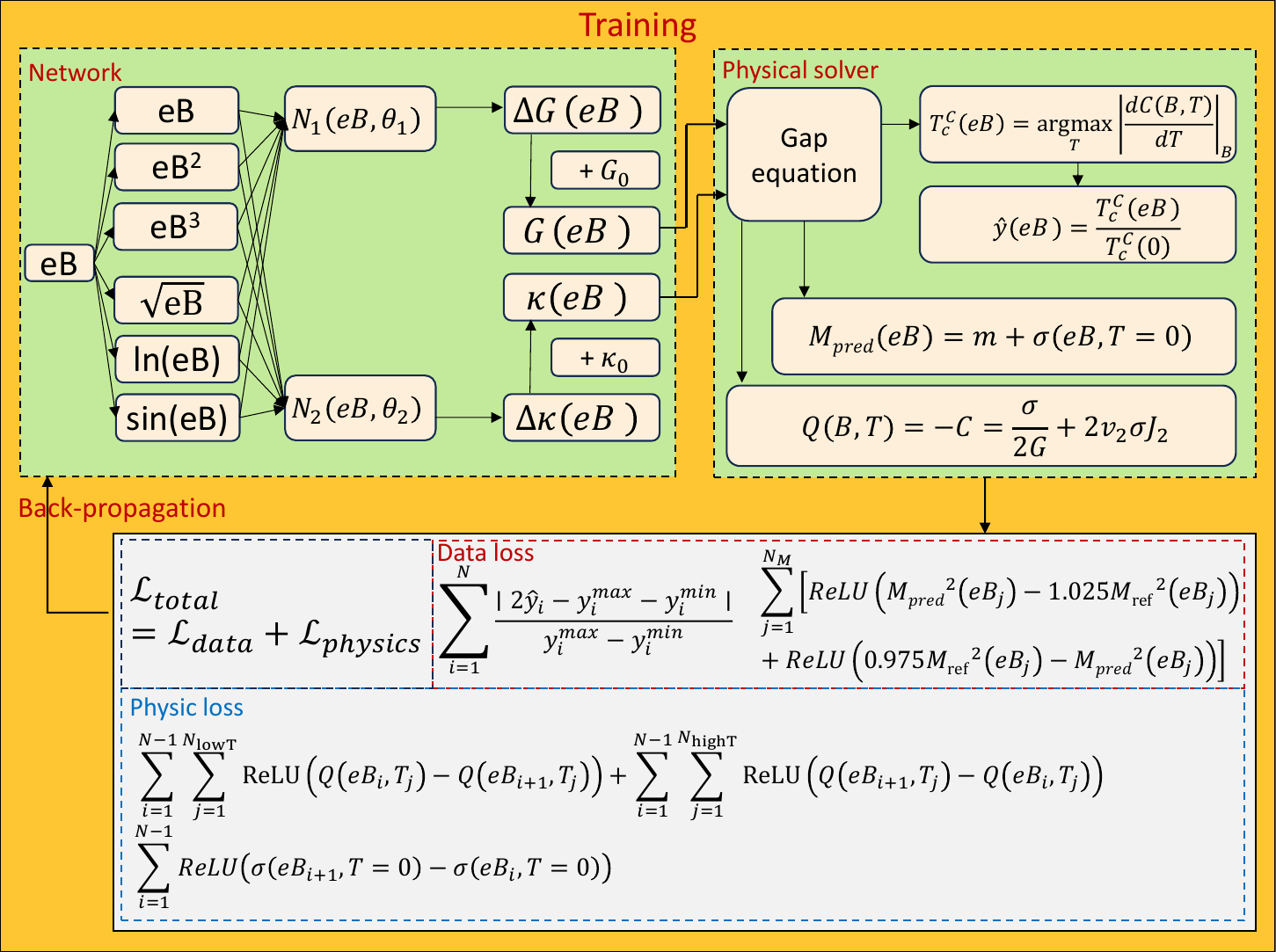}
	\caption{Schematic architecture of the differentiable physics-informed inversion framework for extracting the magnetic field-dependent parameters $G(eB)$ and $v_2(eB)$ from lattice QCD data. The framework integrates a neural network with feature-expanded inputs, the gap equation as a hard physics solver, and a multi-objective loss function that incorporates both data-fidelity and physical-prior constraints. }
	\label{Fig1}
\end{figure}

We developed an inversion framework based on physics-informed learning to extract the magnetic-field-dependent parameters $G(eB)$ and $v_{2}(eB)$ from lattice QCD data. This framework integrates the gap equation (Eq.~\eqref{eq:gap}) as a comprehensive physics solver within the computational workflow and incorporates prior-knowledge constraints to form an end-to-end differentiable computational graph. Optimal parameters are determined by minimizing the following multi-objective loss function
\begin{equation}
	\hat{\theta} = \arg\min_{\theta} \mathcal{L}_{\mathrm{total}}(\theta, \mathcal{D}_{\mathrm{latt}}, \mathcal{P}),
\end{equation}
where $\theta$ denotes the neural network parameters, $\mathcal{D}_{\mathrm{latt}}$ represents the dataset of lattice QCD observables, and $\mathcal{P}$ stands for the encoded physical prior knowledge. As illustrated in Fig.~\ref{Fig1}, feature-expanded $eB$ inputs feed two residual networks for $G$ and $v_2$. Their outputs enter the total-derivative gap solver, which determines $\sigma$ and the derived observables $M=m+\sigma$, $Q=-C$, and $T_c^C(eB)/T_c^C(0)$. The data-loss blocks compare the crossover ratio and $M^2$ with their target ranges; the physics-loss blocks constrain the field dependence of $\sigma$ and the low- and high-temperature trends of $Q$. The weighted objective is back-propagated through the differentiable solver to update both networks.  Crucially, the embedded gap equation acts as a ``hard constraint" to enforce the chosen model stationarity condition, while physics-based regularization terms in the loss function serve as ``soft constraints", guiding the solution toward regions consistent with theoretical expectations.

In Fig.~\ref{Fig1}, $\kappa$, $\Delta\kappa$, and $\kappa_0$ label $v_2$, $\Delta v_2$, and $v_2^0$; the physical AMM is $\kappa=v_2\sigma^2$. Exact input offsets, loss normalizations, margins, auxiliary terms, and weights are specified below.

To achieve flexible modeling of the unknown parameter functions $G(eB)$ and $v_2(eB)$, we employ two fully connected neural networks (FCNNs) with identical architectures, each dedicated to parameterizing one function. Each network consists of six hidden layers with 64, 128, 64, 32, 16, and 16 neurons, respectively, with ReLU activation after the first five hidden layers. Given that the magnetic field strength $eB$ is a single scalar input, we first perform feature expansion to construct a six-dimensional feature vector, enhancing the network's ability to capture physical scale dependencies: $\phi(eB) = \bigl[ eB, \; eB^2, \; eB^3, \; \sqrt{|eB|+\epsilon}, \; \ln(|eB|+\epsilon), \; \sin(eB) \bigr]$, where $\epsilon = 10^{-6}$ ensures numerical stability. This feature vector serves as the network's input, implicitly encoding potential dependency patterns, such as low-order polynomials, logarithmic terms, and periodic behavior. The network outputs are defined as adjustments relative to the reference values  ($G_0 = 4.97$ GeV$^{-2}$, $v_2^0 = 0.8$ GeV$^{-3}$), yielding the final parameter functions: $G(eB) = G_0 + \Delta G(\phi(eB),\theta_1), v_2(eB) = v_2^0 + \Delta v_2(\phi(eB),\theta_2)$, where $\theta_1$ and $\theta_2$ are the trainable parameters of the two networks. This residual parameterization scheme allows the networks to learn only the deviations from fixed baselines, significantly reducing the learning burden and improving training efficiency.

To guarantee the physical validity of the learned parameters, we integrate the gap equation as a hard constraint into the framework. During training, we employ $N = 15$ discrete magnetic field strength values $\{eB_i\}_{i=1}^{N}$ uniformly distributed from $0$ to $0.6~\text{GeV}^2$, serving as fixed inputs for each training epoch. For each parameter set $\left(G(eB_i), v_2(eB_i)\right)$ output by the neural network, we solve the corresponding gap equation numerically to obtain $\sigma(eB_i,T)$. The consistent observables follow from Eqs.~\eqref{eq:mass_field} and \eqref{eq:source_condensate}: the zero-temperature constituent mass $M_{\mathrm{pred}}$, the positive scalar-condensate magnitude $Q\equiv-C$, and the normalized crossover temperature. We define the chiral pseudo-critical temperature by the main peak of the temperature derivative of the mass-source condensate in the crossover region:
\begin{equation}
	\label{eq:Tc_source}	
		T_c(eB)\equiv T_c^C(eB)
		=\operatorname*{arg\,max}_{T}\left|\frac{dC(B,T)}{dT}\right|_{B},
		\qquad
		\hat y(eB_i)=\frac{T_c^C(eB_i)}{T_c^C(0)}.
\end{equation}
The derivative is evaluated along equilibrium gap solutions at fixed $B$, $G(eB)$, and $v_2(eB)$. The temperature scan uses 30 points from $0.005$ to $0.300~\mathrm{GeV}$, with local quadratic refinement of the slope maximum. Implicit differentiation of the gap equation and differentiation through the peak interpolation enable gradient-based optimization on the selected branch.

Using these differentiable observables, we formulate an optimization objective that reconciles the model's predictions with both empirical lattice data and fundamental QCD phenomenology. The total loss function $\mathcal{L}_{\text{total}}(\theta)$ is defined as
\begin{equation}
	\label{eq:total_loss}
	\begin{aligned}
		\mathcal L_{\mathrm{total}}={} & \lambda_b\mathcal L_b+\lambda_M\mathcal L_M
		+\lambda_\sigma\mathcal L_\sigma
		+\lambda_{QC}(\mathcal L_{QC}^{\mathrm{low}}+\mathcal L_{QC}^{\mathrm{high}}).
	\end{aligned}
\end{equation}
where $\lambda_i$ are hyperparameters that balance data fidelity and physical consistency. The individual components are defined as follows:
\begin{enumerate}[leftmargin=*]
	\item Data-Fidelity Terms

	      To ensure the model accurately reproduces the phase boundary and mass scales determined by lattice QCD, we
	      define:
	      \begin{itemize}
		      \item Normalized Critical Temperature Boundary Loss $$\mathcal{L}_{b} = \sum_{i=1}^{N} \frac{|2\hat{y}_i - (y_i^{\max} + y_i^{\min})|}{\Delta y_i},$$ where $\Delta y_i = y_i^{\max} - y_i^{\min}$ is the target interval width. This term provides a continuous gradient to drive the predicted normalized critical temperature $\hat{y}_i$ toward the center of the target range. Here $\hat y_i$ is the source-condensate temperature ratio defined in Eq.~\eqref{eq:Tc_source}.
		      \item Constituent Quark Mass Loss:$$\mathcal{L}_M = \sum_{j=1}^{N_M} \left[ \operatorname{ReLU}\bigl( {M}_{\text{pred}}^2(eB_j) - 1.025 M_{\text{ref}}^2(eB_j) \bigr) + \operatorname{ReLU}\bigl( 0.975 M_{\text{ref}}^2(eB_j) - {M}_{\text{pred}}^2(eB_j) \bigr) \right],$$which penalizes predictions falling outside a $\pm 2.5\%$ tolerance zone of
		            the reference squared masses $M_{\mathrm{ref}}^2$ (see Table~\ref{Table1}), where $N_M = 8$ corresponds to the number of magnetic field values provided in the table.
	      \end{itemize}
	\item Physics-Informed Regularizer

	      To guide the neural network outputs, we introduce the following phenomenological priors:
	      \begin{itemize}
		      \item Chiral Dynamical Quark Mass Monotonicity Loss $$\mathcal{L}_\sigma = \sum_{i=1}^{N-1} \operatorname{ReLU}\bigl( \sigma(eB_{i+1}, T=0) - \sigma(eB_i, T=0) \bigr),$$which encodes the physical behavior that $\sigma$ decreases monotonically with increasing magnetic field at zero
		            temperature.
		      \item Quark Condensate Loss $$\mathcal{L}_{QC}^{\text{low}} = \sum_{i=1}^{N-1} \sum_{j=1}^{N_{\text{lowT}}} \operatorname{ReLU}\bigl( Q(eB_i, T_j) - Q(eB_{i+1}, T_j) \bigr),$$$$\mathcal{L}_{QC}^{\text{high}} = \sum_{i=1}^{N-1} \sum_{j=1}^{N_{\text{highT}}} \operatorname{ReLU}\bigl( Q(eB_{i+1}, T_j) - Q(eB_i, T_j) \bigr),$$which drives the positive magnitude $Q$ to increase with the magnetic field—exhibiting MC on the low temperature set $\{T_j\} = \{0, 0.05\}~\text{GeV}$ ($N_{\text{lowT}} = 2$), and to decrease with the field—showing IMC on the high temperature set $\{T_j\} = \{0.175, 0.24\}~\text{GeV}$ ($N_{\text{highT}} = 2$).
	      \end{itemize}
\end{enumerate}

The weighting coefficients $\lambda_i$ are empirically determined (see Table~\ref{Table2}) to implement a hierarchical optimization priority. We assign significantly higher magnitudes to the physics-informed weights ($\lambda_\sigma, \lambda_Q$) and the constituent quark mass constraint ($\lambda_M$) relative to the temperature weight $\lambda_T$. This disparity reflects the role of physical priors as stringent boundary conditions. Crucially, as these constraints are implemented via the ReLU functional—which provides a gradient only upon violation—the high weights do not disproportionately dominate the global optimization once the model enters the physically feasible region. This ensures that the primary focus of late-stage learning shifts to refining the accuracy of the phase boundary using $\mathcal{L}_{b}$, without compromising physical validity.

To ensure a robust convergence that honors both empirical observations and theoretical requirements, the model is trained using a two-stage optimization strategy. In the pre-training stage (typically 500 epochs), the network is optimized solely with respect to the data-driven terms $\lambda_b\mathcal{L}_{b} + \lambda_M \mathcal{L}_M$. This initial phase allows the model to capture the foundational parameter dependencies from the lattice data without being prematurely constrained by complex physics priors. Subsequently, in the joint optimization stage, the full loss function $\mathcal{L}_{\text{total}}$ is activated to steer the solution toward physically consistent regions while refining the numerical fit. This stage is maintained for at least 1500 epochs and extended until the total loss plateaus, ensuring that the interplay between the high-weight physical constraints and the data-fidelity term has reached a global minimum. The framework is optimized using the Adam optimizer with an initial learning rate of $1 \times 10^{-4}$ and a batch size of 15, effectively performing full-batch gradient descent. To facilitate precise convergence, a ReduceLROnPlateau scheduler (patience: 5, factor: 0.9) is employed, allowing the learning rate to adaptively diminish whenever the optimization progress stagnates across both training phases.

\begin{table*}
	\centering
	\begin{ruledtabular}
		\begin{tabular}{lccc}
			Category       & Hyperparameter                                  & Symbol                 & Value             \\
			\colrule
			Data Fidelity  & Normalized critical temperature boundary weight & $\lambda_b$            & 1.0               \\
			               & Constituent quark mass weight                   & $\lambda_M$            & $1.5\times10^{3}$ \\
			\colrule
			Physics Priors & Chiral dynamical quark mass monotonicity weight & $\lambda_\sigma$       & $10^5$            \\
			               & Quark condensate weight                         & $\lambda_{QC}$         & $10^4$            \\
			\colrule
			Optimization   & Initial learning rate                           & $\eta_{\mathrm{init}}$ & $1\times10^{-4}$  \\
		\end{tabular}
	\end{ruledtabular}
	\caption{Configuration of hyperparameters and weighting coefficients for the multi-objective loss function.}
	\label{Table2}
\end{table*}

\section{Numerical Results }

\label{sec:results}
\begin{figure}[t]
	\centering
	\includegraphics[width=0.45\textwidth]{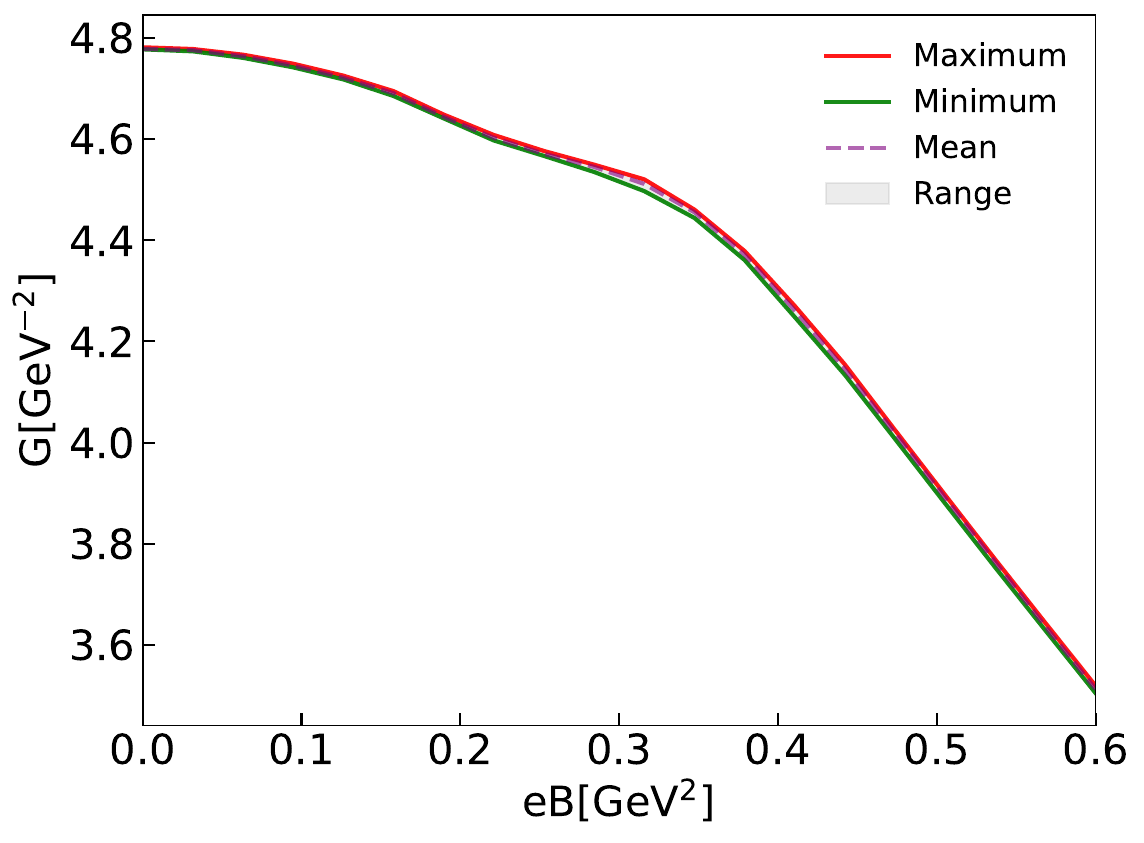}
	\includegraphics[width=0.45\textwidth]{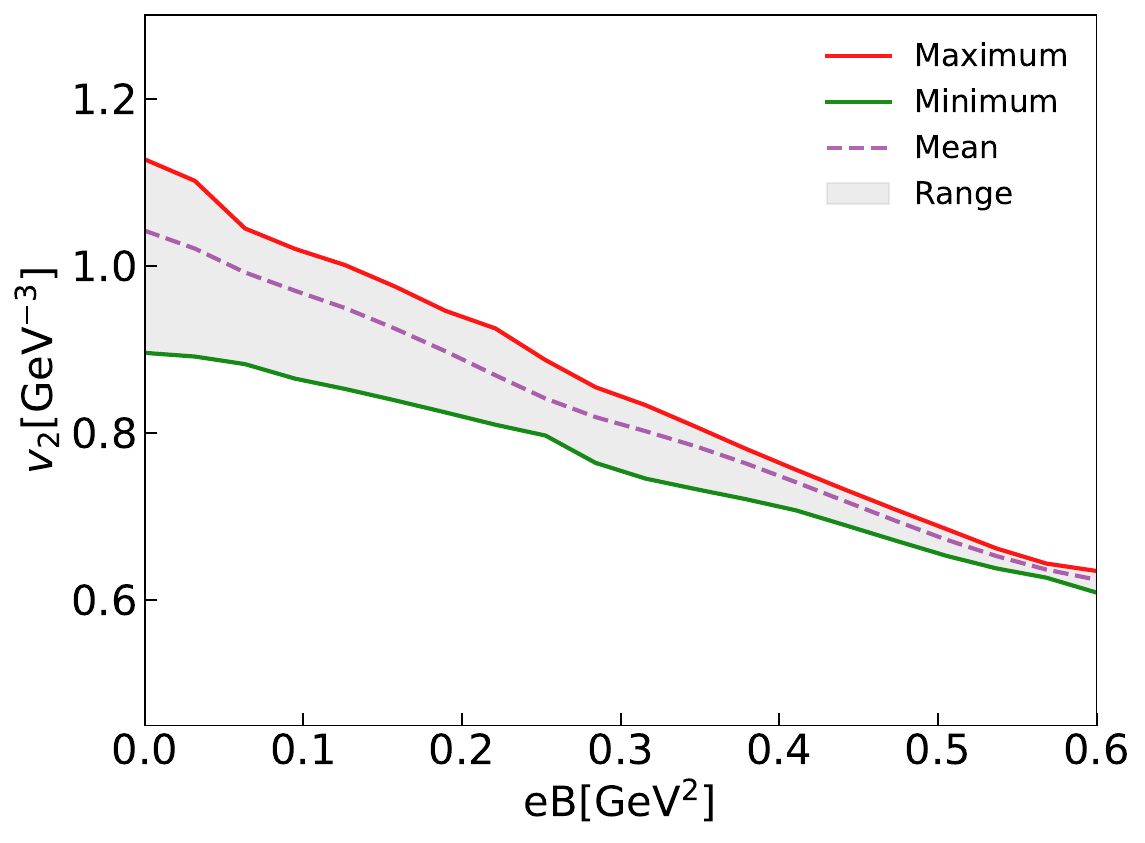}
	\caption{Estimated value ranges for the coupling constant $G$ (left) and the AMM ratio $v_{2}$ (right) as functions of the magnetic field $eB$. The shaded regions indicate the spread of values obtained across repeated training runs. }
	\label{FigZone}
\end{figure}

\begin{figure}[t]
	\centering
	\includegraphics[width=0.45\textwidth]{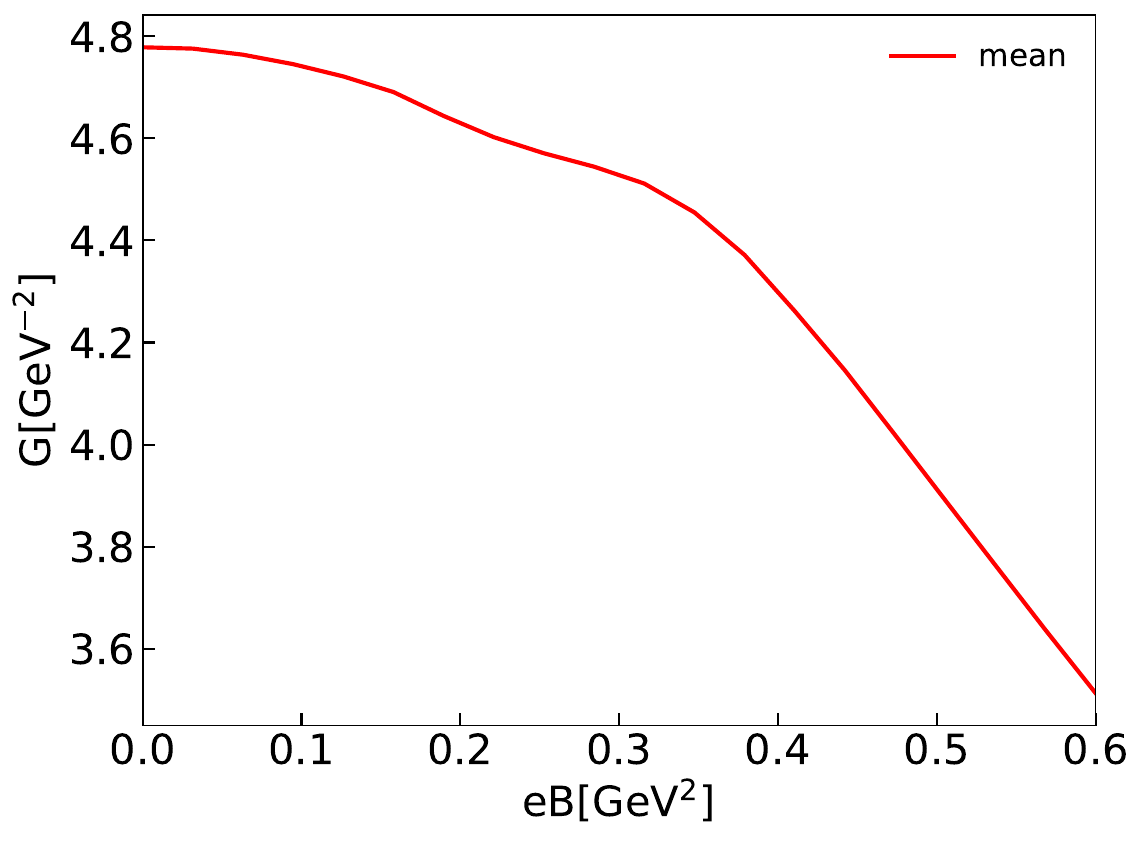}
	\includegraphics[width=0.45\textwidth]{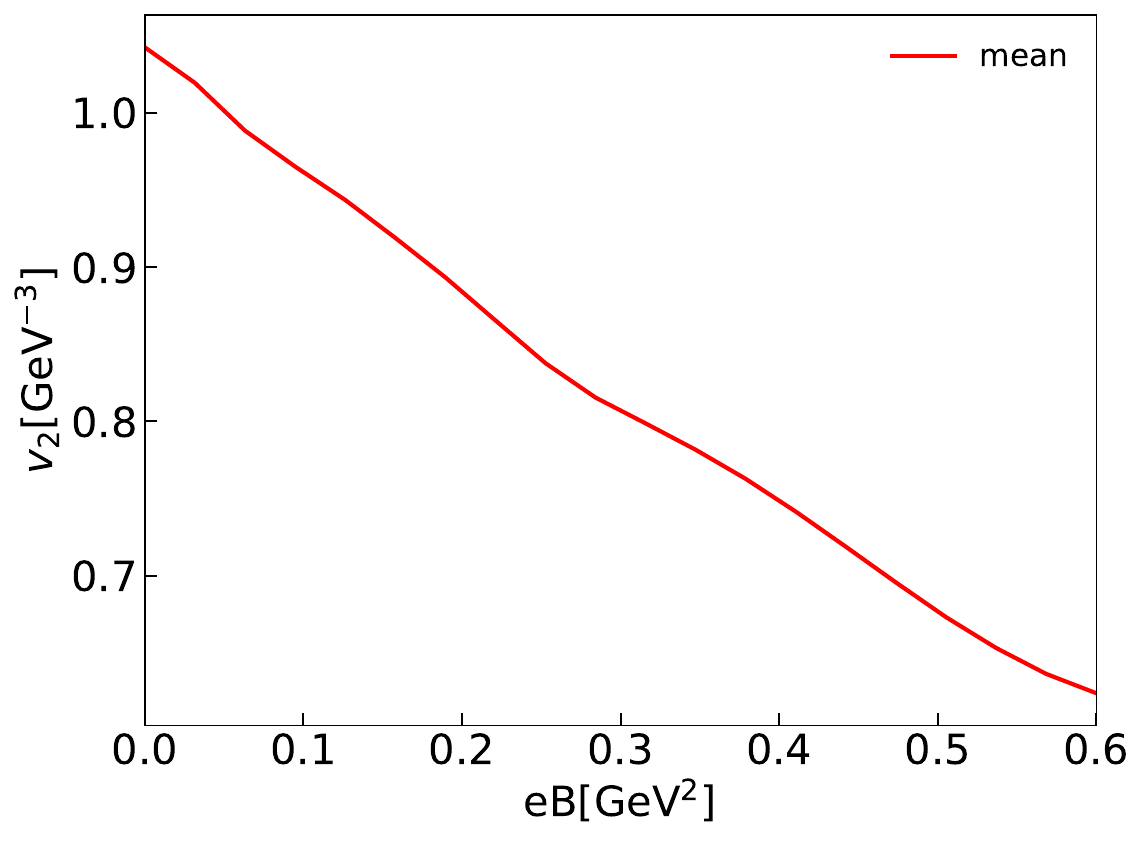}
	\caption{The mean value of the coupling constant $G$(left) and the AMM constant $v_2$(right) as a function of $eB$. }
	\label{FigGsv2eB}
\end{figure}

\begin{table*}
	\begin{ruledtabular}
		\begin{tabular}{lcccccccc}
			$\text{eB}~[\text{GeV}^2]$     & 0.0 & 0.1 & 0.2 & 0.3 & 0.4 & 0.5 & 0.6 \\
			\colrule
			$\text{G}~[\text{GeV}^{-2}]$   &
			$4.778 \pm 0.002$              &
			$4.741 \pm 0.002$              &
			$4.629 \pm 0.003$              &
			$4.530 \pm 0.005$              &
			$4.298 \pm 0.006$              &
			$3.913 \pm 0.005$              &
			$3.513 \pm 0.005$                                                        \\
			\colrule
			$\text{v}_2~[\text{GeV}^{-3}]$ &
			$1.043 \pm 0.076$              &
			$0.962 \pm 0.052$              &
			$0.884 \pm 0.036$              &
			$0.807 \pm 0.027$              &
			$0.749 \pm 0.015$              &
			$0.677 \pm 0.009$              &
			$0.624 \pm 0.008$                                                        \\
		\end{tabular}
	\end{ruledtabular}
	\caption{Mean values and corresponding standard deviations of the coupling constant $G$ and the AMM constant $v_2$ at selected key magnetic field strengths.}
	\label{Table3}
\end{table*}

By reconciling the NJL model with lattice QCD observations, our physics-informed machine learning framework extracts the magnetic-field dependence of both the coupling constant $G(eB)$ and the AMM parameter $v_{2}(eB)$. The continuous evolution of these parameters, evaluated over 100 independent training runs, is illustrated in Fig.~\ref{FigZone}, with specific values at key field strengths summarized in Table~\ref{Table3}.

Within the investigated range of $0 < |eB| < 0.6 \text{ GeV}^2$, we observe that the magnetic field smoothly suppresses both parameters. Specifically, the coupling constant $G(eB)$ decreases from a vacuum value of $4.778 \pm 0.002~\text{ GeV}^{-2}$ to $3.513 \pm 0.005~\text{GeV}^{-2}$ at the highest field strength. The high consistency across training runs (relative uncertainty $< 0.2\%$) suggests that the suppression of $G(eB)$ is a robust requirement for reproducing the lattice-determined chiral transition. This downward trend physically represents the screening of the strong interaction in the presence of an intense magnetic background.

Similarly, the AMM ratio $v_{2}$ is reduced from $1.043 \pm 0.076~\text{GeV}^{-3}$ to $0.624 \pm 0.008~\text{GeV}^{-3}$. While $v_{2}$ exhibits a larger relative uncertainty (approximately $5\%$), its overall suppression confirms that the relationship between the anomalous magnetic moment and the chiral condensate is not a simple constant but is modulated by the external field. Furthermore, our analysis confirms that within this framework, the magnitude of $v_{2}$ approximately linearly suppressed by the magnetic field in this region. The evolution of the mean parameter values is depicted in  Fig.~\ref{FigGsv2eB}.

\begin{figure}[t]
	\centering
	\includegraphics[width=0.45\textwidth]{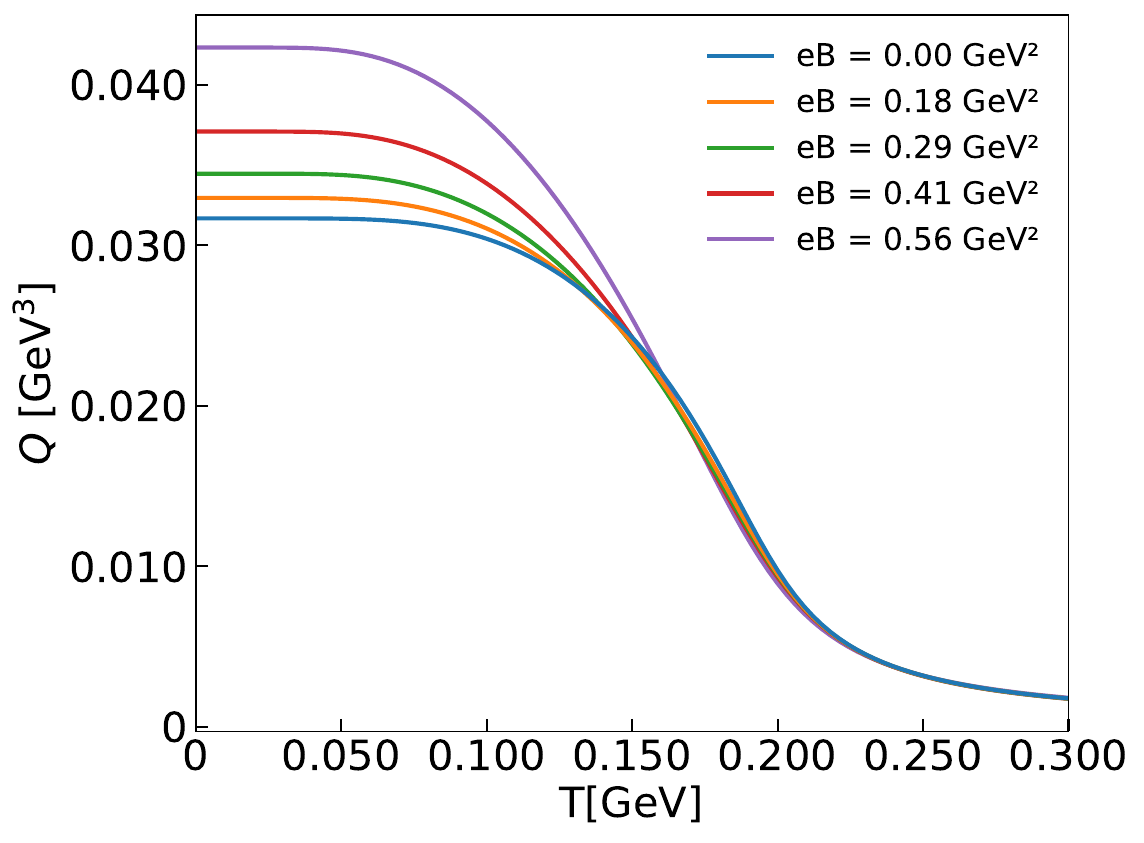}
	\includegraphics[width=0.45\textwidth]{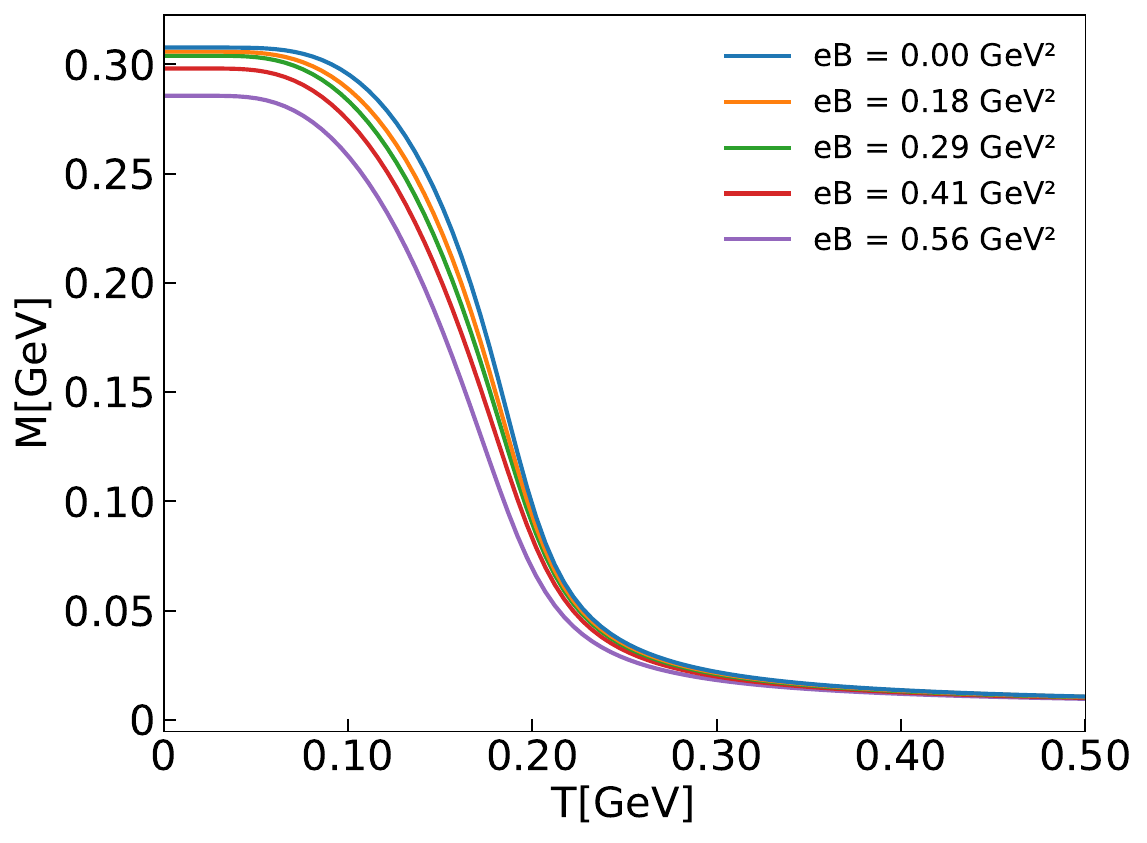}
	\caption{The mass-source condensate magnitude $Q$ and constituent mass of quarks with the mean value of our running parameters. }
	\label{FigMT}
\end{figure}

Using the mean values of these learned parameters, we computed the temperature-dependent mass-source condensate magnitude $Q$ and constituent-quark mass for several magnetic-field strengths, as shown in Fig.~\ref{FigMT}. The chiral condensate displays a complex, temperature-dependent dual behavior: at low temperatures, the magnetic catalysis effect dominates as the field enhances the condensate; however, near the transition region, the field-dependent parameters trigger a suppression of the condensate, successfully capturing the inverse magnetic catalysis effect observed in lattice QCD. The condensate enhancement and constituent-mass reduction are compatible because the gap equation gives $Q=\sigma/(2G)+2v_2\sigma J_2$, rather than $Q=\sigma/(2G)$.

\begin{figure}[t]
	\centering
	\includegraphics[width=0.45\textwidth]{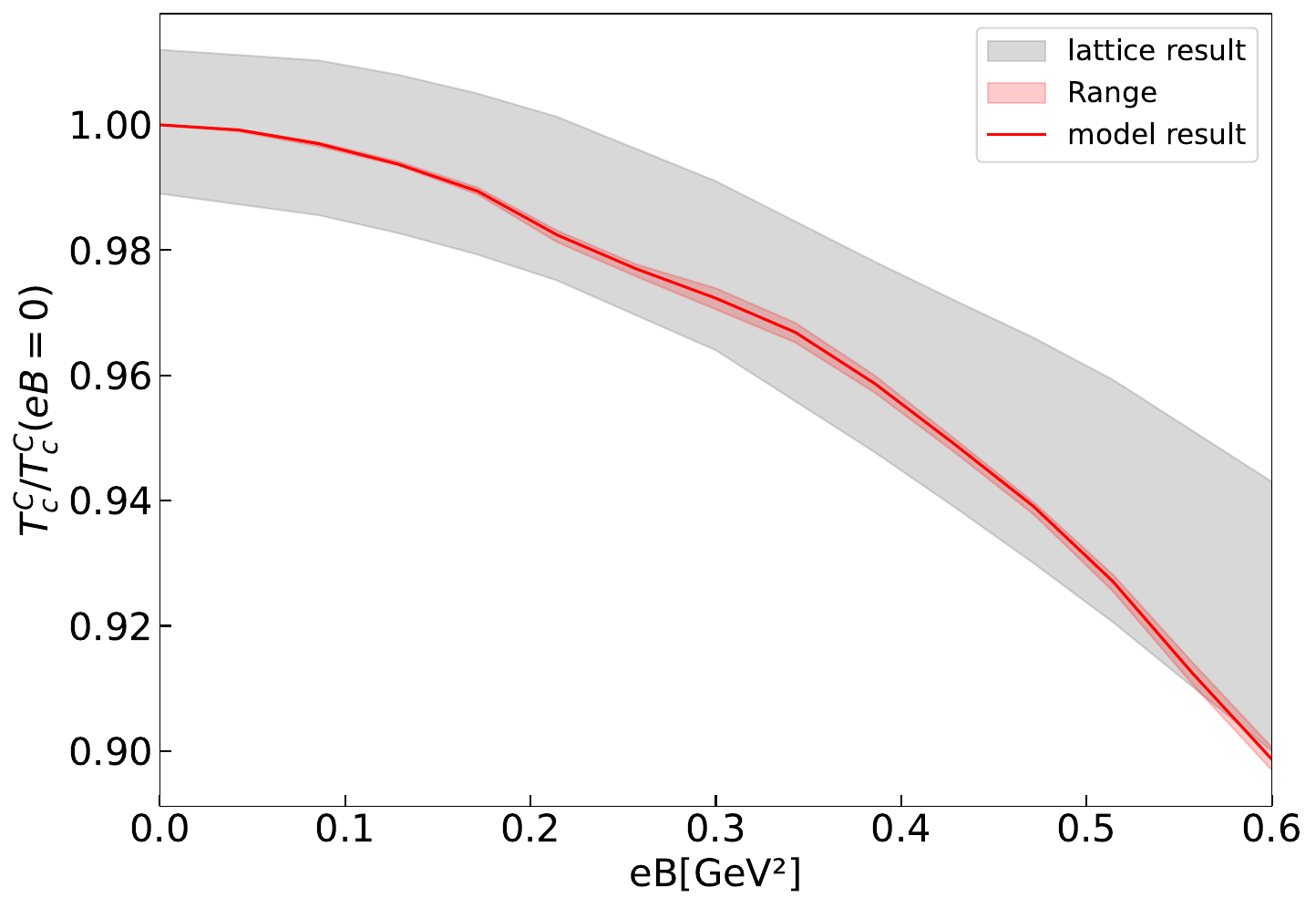}
	\caption{The normalized pseudo-critical temperature $T_c^C(eB)/T_c^C(0)$, as a function of the magnetic field. The results from the physics-informed framework (red shade and red line for mean value) are compared against lattice QCD data (gray shade).}
	\label{FigTC}
\end{figure}

The constituent quark mass $M$ decreases monotonically with increasing $eB$ across the entire studied temperature range. This behavior aligns with the learned reduction in the coupling strength and is consistent with previous theoretical studies~\cite{Sheng:2021evj} that identify mass reduction as a precursor to chiral restoration. Finally, the normalized pseudo-critical temperature $T_c(eB)/T_c(eB=0)$ is plotted against lattice constraints~\cite{Bali:2012zg} in Fig.~\ref{FigTC}. Our framework achieves excellent agreement with the lattice data, with a maximum discrepancy of less than $1\%$. This result confirms that the machine-learned functional forms for $G(eB)$ and $v_2(eB)$ provide a highly accurate parameterization for the NJL model in extreme magnetic environments. However, since this quantity is used in training, an independent test is presented in Sec.~\ref{sec:pion_results}.

\FloatBarrier
\section{Independent validation with the neutral-pion mass}
\label{sec:pion_results}
\suppressfloats[t]

To examine the predictive capability and generalization power of our machine-learning framework beyond the training observables, we calculate the neutral-pion pole mass in the vacuum ($T=\mu=0$) using the mean running parameters $\overline{G}(eB)$ and $\overline{v}_2(eB)$ obtained from repeated optimization runs. Crucially, neutral-pion observables were excluded entirely from the training loss function, and all model parameters remain fixed without any post-hoc adjustments or secondary refitting.

The neutral-pion pole is extracted within the random-phase approximation (RPA)~\cite{Avancini:2016pion,Sheng:2021polescreen}. Minimal electromagnetic coupling preserves the neutral axial direction in the chiral limit. However, the scalar Pauli AMM term in Eq.~(1) is not by itself invariant under neutral axial rotations~\cite{Chang:2010hb}. The background relation $\kappa(\sigma)=v_2\sigma^2$ specifies the scalar AMM but does not determine the action for pseudoscalar fluctuations. On the spontaneously broken branch ($\sigma>0$), we therefore adopt the neutral-axial-covariant completion
\begin{equation}
	\label{eq:pion_completion}
	\begin{gathered}
			\rho=\sqrt{\sigma^2+\pi^2},\qquad \Phi_5=\sigma+i\gamma_5\tau_3\pi,\\
			\mathcal M=m+\Phi_5-v_2\rho\mathcal A_B\Phi_5,\qquad U=\frac{\rho^2}{4G},\\
			\mathcal A_B=\frac12q_fF_{\mu\nu}\sigma^{\mu\nu}.
		\end{gathered}
\end{equation}
Here $U$ is the bosonic potential. For a purely magnetic background and $m=0$, this completion is invariant under simultaneous neutral axial rotations of the quark and $(\sigma,\pi)$ fields. In the limit of vanishing pion fluctuations, this construction smoothly recovers the constituent quark mass $M=m+\sigma$, the original AMM ansatz, and the mean-field gap equation. Expanding the single-quark Hamiltonian in powers of the pion field, $H = H_0 + \pi V_\pi + \frac{1}{2}\pi^2 W_{\pi\pi} + \cdots$, naturally generates both an AMM-dressed pion vertex and a contact interaction:
\begin{equation}
	V_\pi=i\gamma^0\gamma_5\tau_3(1-v_2\sigma\mathcal A_B),
		\qquad W_{\pi\pi}=-v_2\gamma^0\mathcal A_B.
\end{equation}
Retaining both contributions is essential for variational consistency. In earlier studies (e.g., Ref.~\cite{Lin:2022ied}), the pion polarization function was evaluated using bare pseudoscalar vertices ($\gamma^0\gamma_5\tau_3$) with AMM-dressed propagators. Within the present completion, however, omitting the induced vertex modification and the contact term would violate the Goldstone theorem, as the RPA kernel would no longer coincide with the exact second functional variation $\delta^2\Gamma/\delta\pi^2$ of the effective action. Incorporating both $V_\pi$ and $W_{\pi\pi}$ guarantees that the Ward-Takahashi identity is preserved and the neutral pion emerges as a genuine pseudo-Goldstone boson satisfying the Gell-Mann--Oakes--Renner relation.

Let $|i\rangle$ and $\epsilon_i$ denote the eigenstates and energy eigenvalues of $H_0$ within each flavor and Landau sector, and define the transition energies $\Delta_{ij} \equiv \epsilon_j - \epsilon_i$. At vanishing spatial momentum ($\mathbf{q}=\mathbf{0}$), defining $z = q_0^2$, the inverse pion propagator kernel is given by
\begin{equation}
	\label{eq:pion_pole}
	\begin{aligned}
			K_\pi(z)           & =\frac1{2G}+\sum_{f,l}\int d\mu_{fl}
			\left[\sum_{i\in-}\langle i|W_{\pi\pi}|i\rangle
			-\sum_{\substack{i\in-                                                   \\j\in+}}
			\frac{2\Delta_{ij}|\langle i|V_\pi|j\rangle|^2}{\Delta_{ij}^2-z}\right], \\
			K_\pi(m_{\pi^0}^2) & =0,\qquad
			d\mu_{fl}=\frac{N_c|q_fB|}{4\pi^2}\,dp_z\,f_{\Lambda,q_fB}^{p_z}.
		\end{aligned}
\end{equation}
Here, $-$ and $+$ label occupied negative-energy sea states and unoccupied positive-energy particle states, respectively. In accordance with the Landau quantization rules, the lowest Landau level ($l=0$) hosts a single physical spin state ($s=\operatorname{sign}(q_f)$), whereas all higher levels ($l \ge 1$) sum over both spin orientations ($s=\pm 1$). Each momentum loop integral incorporates the identical soft-cutoff regulator $f_{\Lambda,q_fB}^{p_z}$ employed in the gap equation, ensuring strict regularitative consistency between the one-body and two-body sectors (reducing to the standard three-dimensional momentum cutoff at $B=0$). The physical neutral-pion mass $m_{\pi^0}$ is then identified as the isolated real pole located below the continuous quark-antiquark threshold.

\begin{figure}[!htb]
	\centering
	\includegraphics[width=0.48\textwidth]{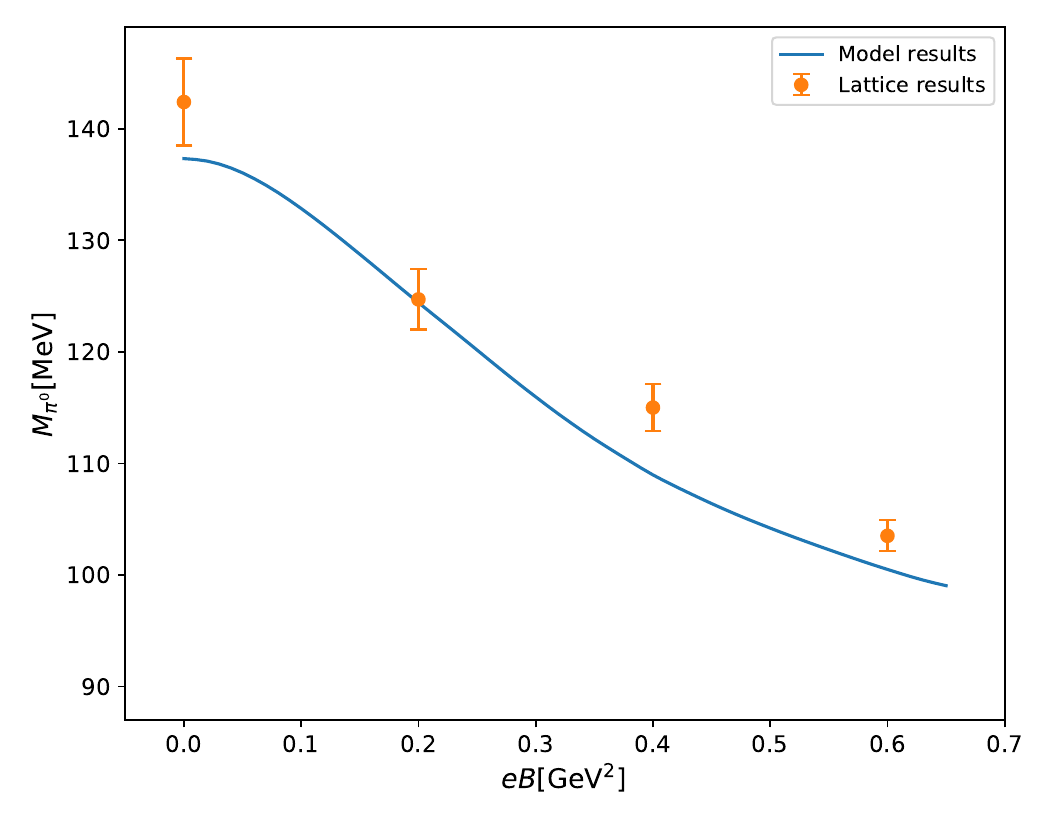}
	\includegraphics[width=0.48\textwidth]{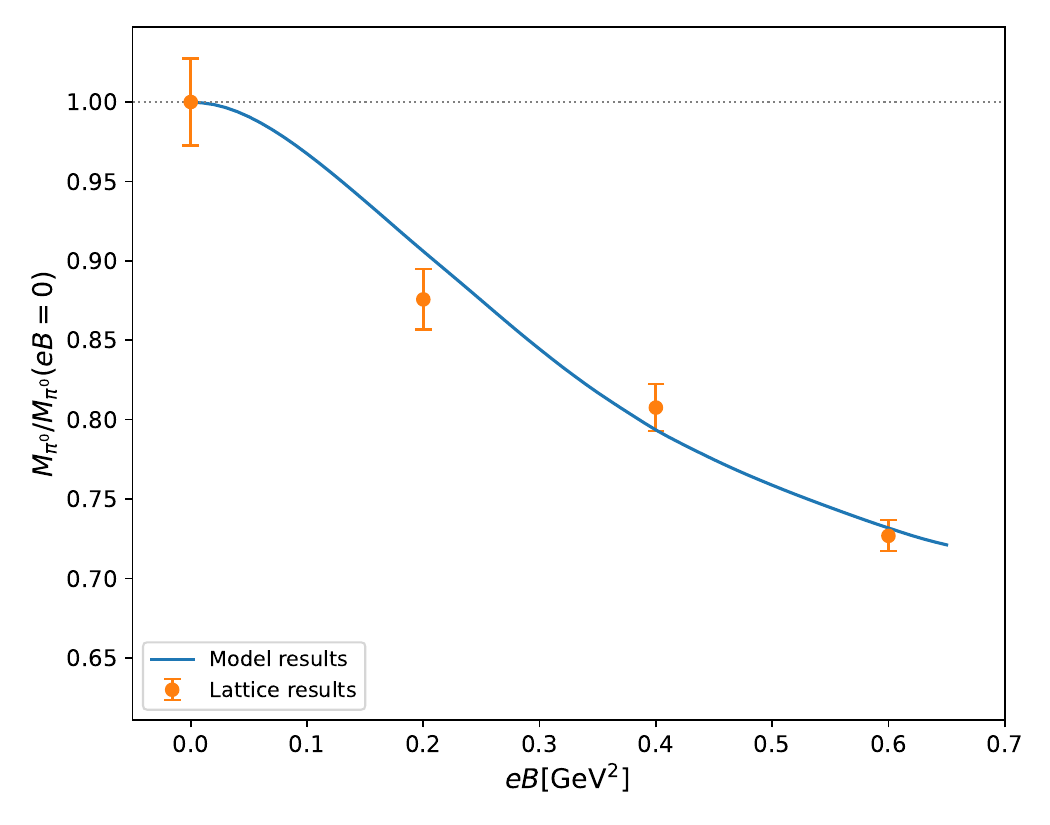}
	\caption{Neutral-pion pole mass $m_{\pi^0}$ (left) and its self-normalized ratio $m_{\pi^0}(eB)/m_{\pi^0}(0)$ (right) as functions of the magnetic field strength. Solid curves represent model predictions evaluated using the mean learned parameters $\overline{G}(eB)$ and $\overline{v}_2(eB)$ without refitting to meson observables, compared against continuum-extrapolated lattice QCD results from Ding and Zhang~\cite{Ding:2026chiral}.  Error bars on the lattice mass ratios reflect the original uncertainties scaled by the central zero-field mass (with $m_{\pi^0}(0)/m_{\pi^0}(0) \equiv 1$). Theoretical curves are extended up to $0.65~\mathrm{GeV}^2$ to illustrate the qualitative trend beyond the $0.6~\mathrm{GeV}^2$ training interval.}
	\label{FigPion}
\end{figure}

In Fig.~\ref{FigPion}, we compare the model predictions with the continuum-extrapolated, connected neutral-pion masses computed in lattice QCD by Ding and Zhang~\cite{Ding:2026chiral}. To facilitate a direct shape comparison that is independent of absolute scale calibration, the right panel displays each dataset normalized to its respective zero-field value. The model successfully reproduces the monotonic decrease of the neutral-pion mass with increasing magnetic field, demonstrating good qualitative and semi-quantitative agreement with lattice observations. Because pion data were entirely absent from the network training objective, this non-trivial consistency across magnetic fields provides strong independent support for the physical validity of the learned running parameters $\overline{G}(eB)$ and $\overline{v}_2(eB)$, confirming that the inverse magnetic catalysis captured in the vacuum condensate naturally extends to the pseudoscalar spectrum.

Several theoretical considerations provide context for this comparison. First, the theoretical curves are evaluated directly at the mean learned parameters $(\overline{G}, \overline{v}_2)$ rather than through an ensemble average over individual training runs, which accounts for the absence of a propagated parameter error band. Second, at strong magnetic fields ($eB \gtrsim \Lambda^2 \approx 0.41~\mathrm{GeV}^2$), the magnetic energy scale approaches and surpasses the three-dimensional momentum cutoff, rendering quantitative predictions increasingly sensitive to the regulator profile. Nevertheless, across the primary training interval, the consistent reproduction of the lattice trend confirms the predictive utility and physical viability of the extracted running parameters.

\FloatBarrier
\section{Conclusion and Outlook}
\label{sec:conclusion}
In this study, we have successfully integrated a physics-informed machine learning pipeline with the NJL model to investigate magnetized QCD matter. Using lattice QCD computation on the chiral transition and condensate behavior, we extracted the optimal functional forms for the running coupling constant $G(eB)$ and the anomalous magnetic moment ratio $v_2(eB)$. Our findings demonstrate that both $G$ and $v_2$ are smoothly suppressed by the external magnetic field. This learned field dependence allows the NJL model to naturally resolve the long-standing discrepancy regarding the inverse magnetic catalysis effect at high temperatures. We also estimate the quark AMM as $\kappa=v_2(eB)\sigma^2$, using the learned mean parameters and the corresponding gap-equation solution. The condensate is defined by its mass source, while the constituent mass remains $M=m+\sigma$. The additional neutral-pion calculation reproduces the decreasing lattice trend without refitting the learned parameters, providing an independent test of their predictive usefulness. In the end, it is important to note that the reliability of the model may decrease in the regime where $|eB| > \Lambda^2 \approx 0.4058 \text{ GeV}^2$, as the field strength begins to exceed the ultraviolet cutoff scale.

This differentiable inversion framework establishes a general paradigm for bridging the gap between first-principles calculations and phenomenological models. Further tests with additional observables and propagated parameter uncertainties will strengthen this assessment. Future work will focus on the following areas: Firstly, extending the pipeline to more sophisticated effective theories, such as the Polyakov-loop NJL (PNJL) model or (2+1)-flavor systems, to provide a more comprehensive description of the QCD transition region. Secondly, utilizing the extracted running parameters to investigate the QCD phase diagram at finite baryon density, which may help identify the location of the critical end point under extreme magnetic conditions. Finally, this inverse-engineering approach can be applied to other challenging regimes in nuclear and particle physics where first-principles data are available but the underlying effective parameters remain poorly constrained. This methodology demonstrates that machine learning is not merely a black-box tool for classification, but a powerful engine for discovery and parameter optimization within the context of fundamental physical laws.

\acknowledgements
The authors thank Zhongmin Niu for insightful discussions and Minghua Wei for early contributions to this work. F.L. acknowledges support from the National Natural Science Foundation of China (Grant No. 12547138) and the Anhui University of Science and Technology (Grant No. 2025yjrc0143). The work of X. W. was supported by the NSFC under Grants No. 12675170 and No. 12635010, and by the Anhui University of Science and Technology under Grant No. YJ20240001.

\bibliographystyle{unsrt}
\bibliography{references}

\end{document}